%% file: duality_polarization_OAM_path_final.tex
\documentclass[pra,aps,showpacs,reprint,twocolumn,superscriptaddress,amsmath,amssymb,floatfix,10pt]{revtex4-1}

\usepackage{bbm}

\usepackage{multirow}
\usepackage{graphicx,dcolumn,bm}
\usepackage{mathrsfs}
\usepackage[usenames,dvipsnames]{xcolor}
\usepackage[colorlinks=true,citecolor=MidnightBlue,linkcolor=MidnightBlue,urlcolor=MidnightBlue]{hyperref}
\usepackage[sort&compress]{natbib}
\usepackage{braket}
\usepackage{bbold}
\usepackage{tikz}
\usetikzlibrary{decorations.markings}

\usepackage[framemethod=TikZ]{mdframed}
\usetikzlibrary{decorations}
\usetikzlibrary{shapes,arrows}

\definecolor{blue1}{rgb}{0.72,0.82,0.98}
\definecolor{silver}{rgb}{0.6,0.6,0.6}
\definecolor{silver1}{rgb}{0.8,0.8,0.8}

\renewcommand{\vec}[1]{\mathbf{#1}}

\newcommand{\pbs}[3]{\draw
[opacity=0.9,color=black!50,shading=axis,shading angle=-#3,left
color=black!5,right color=black!20,rotate around={#3:(#1,#2)}]
(#1-0.25,#2-0.25)--++(0.5,0.5)--++(-0.5,0)--++(0,-0.5)--++(0.5,0)--++(0,0.5)}

\newcommand{\filledboxtwo}[4]{\draw [thick,rounded corners,fill=lightgray]
(#1,#2) rectangle (#3,#4)}

\newcommand{\hologram}[3]{
\draw[rotate around={#3:(#1,#2)},line width=1pt] (#1,#2) to(#1+0.25,#2);
\draw[rotate around={#3:(#1,#2)},line width=1.2pt] (#1+0.25,#2-0.075) to(#1+0.25,#2+0.075);
\draw[rotate around={#3:(#1,#2)},line width=1pt, dashed, color=red] (#1,#2) to (#1+0.6,#2+0.6);
\draw[rotate around={#3:(#1,#2)},line width=1pt, dashed, color=red] (#1,#2) to (#1+0.6,#2-0.6);
\draw[rotate around={#3:(#1,#2)},fill=gray] (#1-0.025,#2-0.3) rectangle ++(0.05,0.6)
}

\begin{document}

\title{Entanglement of Polarization and Orbital Angular Momentum}

\author{D. Bhatti}
\affiliation{Institut f\"ur Optik, Information und Photonik, Universit\"at Erlangen-N\"urnberg, 91058 Erlangen, Germany}
\affiliation{Department of Physics, Oklahoma State University, Stillwater, OK, USA}

\author{J. von Zanthier}
\affiliation{Institut f\"ur Optik, Information und Photonik, Universit\"at Erlangen-N\"urnberg, 91058 Erlangen, Germany}
\affiliation{Erlangen Graduate School in Advanced Optical Technologies (SAOT), Universit\"at Erlangen-N\"urnberg, 91052 Erlangen, Germany}

\author{G.~S.~Agarwal}
\affiliation{Department of Physics, Oklahoma State University, Stillwater, OK, USA}

\date{\today}

\begin{abstract}
We investigate two-photon entangled states using two important degrees of freedom of the electromagnetic field, namely orbital angular momentum (OAM) and spin angular momentum. For photons propagating in the same direction we apply the idea of \textit{entanglement duality} and develop schemes to do \textit{entanglement sorting} based either on OAM or polarization. In each case the entanglement is tested using appropriate witnesses. We finally present generalizations of these ideas to three- and four-photon entangled states.
\end{abstract}


\maketitle

\section{Introduction}

It is known that for identical particles one can construct entangled states by using different degrees of freedom, e.g., linear momentum, polarization or even orbital angular momentum (OAM). For instance, for two photons traveling in different directions $\vec{k}_{1}$ and $\vec{k}_{2}$ one can consider an entangled state involving linear momentum and polarization degrees of freedom 
\begin{equation}
	\ket{\Psi_{Pol}}=\frac{1}{\sqrt{2}}\left(\ket{H,\vec{k}_{1}}\ket{V,\vec{k}_{2}}+\ket{V,\vec{k}_{1}}\ket{H,\vec{k}_{2}}\right) \ .
\label{eq:dualHVpath1}
\end{equation}
Such states are now routinely produced by type II parametric down conversion \cite{Shih(1995)} and have been extensively studied in the literature \cite{Zeilinger(2012)}. More recently it was realized that orbital angular momentum is another degree of freedom of the radiation field which can be fruitfully employed for entanglement generation. This led among others to the production of entangled states of the form
\begin{equation}
	\ket{\Psi_{OAM}}=\frac{1}{\sqrt{2}}\left(\ket{l,\vec{k}_{1}}\ket{-l,\vec{k}_{2}}+\ket{-l,\vec{k}_{1}}\ket{l,\vec{k}_{2}}\right) \ ,
\label{eq:dualOAMpath1}
\end{equation}
which have been used in tests of nonlocality as well as for applications in quantum communication and cryptography \cite{Mair(2001),Franke-Arnold(2008),Okamoto(2008),Leach(2009),Karimi(2010),Leach(2010)}.\\
A more interesting possibility would be to consider two photons with the same linear momentum but with different polarizations and OAM degrees of freedom
\begin{equation}
	\ket{\Psi}=\frac{1}{\sqrt{2}}\left(\ket{H,l}\ket{V,-l}+\ket{H,-l}\ket{V,l}\right) \ .
\label{eq:dualPolOAM1}
\end{equation}
In this case we have entanglement between two types of angular momenta, namely spin angular momentum and orbital angular momentum. Entanglement of these two degrees of freedom has not been studied extensively in the literature so far.\\
In this paper we focus our attention precisely on entangled states of the form of Eq.~(\ref{eq:dualPolOAM1}). We show how to produce these entangled states and use the recently formulated \textit{duality} of identical particle entanglement \cite{Bose(2013)} to perform \textit{entanglement sorting}. Note that the state given in Eq.~(\ref{eq:dualPolOAM1}) has the interesting property that one can detect the entangled character by studying either polarization variables or OAM variables. These studies would yield identical information if the particles are indistinguishable.\\
The paper is organized as follows. In Sec.~\ref{sec:2} we describe the production of single path two-photon states displaying polarization-OAM entanglement. To this aim we use different types of polarization to OAM transferrers to generate polarization-entangled photon pairs in separate path modes whereupon a polarizing beam splitter (PBS) is employed to project both photons into a single path. Using entanglement duality we then describe entanglement sorting in sections~\ref{sec:3} and \ref{sec:4}. More specifically, we describe in Sec.~\ref{sec:3} the detection of entanglement in OAM variables, whereas in Sec.~\ref{sec:4} we register the entanglement in the polarization degrees of freedom. In both cases entanglement witnesses are constructed. In Sec.~\ref{sec:5} we generalize the idea of entanglement sorting to three- and four-photon entangled states, and in Sec.~\ref{sec:6} we finally conclude.

\section{Entangled States of Polarization and Orbital Angular Momentum}
\label{sec:2}

\begin{figure}[htpb]
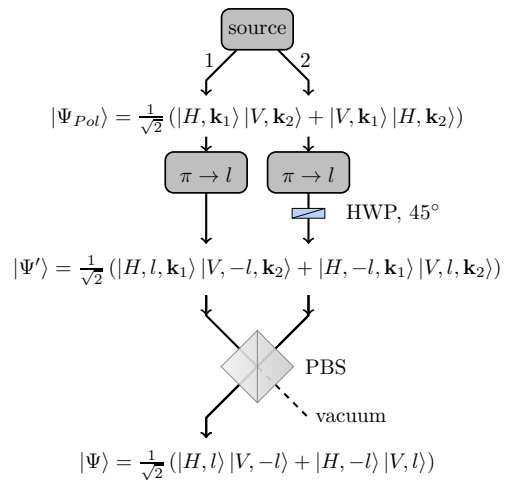

\centering
\scalebox{0.8}{
\include{source2small}}
\caption{A two-photon source is used producing a polarization-entangled state in two spatial modes $1$ and $2$. Polarization to OAM ($\pi\rightarrow l$) transferrers, here consisting of q-plates, transfer horizontal and vertical polarization into OAM $\pm l$, while a second possible $\pi\rightarrow l$ transferrer consists of spatial light modulators (see text for details). A half wave plate (HWP) in path mode $2$ changes the polarization from $H$ to $V$ what leads to a simultaneous entanglement of polarization, OAM, and path. Mixing the two spatial modes through a polarizing beam splitter (PBS) then creates the aspired polarization-OAM entanglement in one single mode.}
\label{fig:source2}
\end{figure}

\begin{figure}[b]
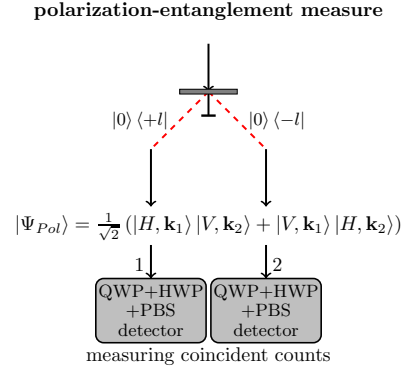

\centering
\scalebox{0.75}{
\include{Polmeasure_new}}
\caption{The polarization-entangled state $\ket{\Psi_{Pol}}$ is generated from the dual polarization-OAM state (cf. Eq.~(\ref{eq:statePolOAM})). A fork hologram is used to diffract the photons and change their OAM by $\Delta l =\pm l$ depending on the direction of diffraction. The dashed (red) lines denote single mode fibers ensuring only detections of photons with OAM changed to $l=0$. The polarization-entanglement is then being verified performing different state projections and coincidence measurements, which give rise to an entanglement witness (cf. Eq.~(\ref{eq:WitnessPol})). Different settings of a quarter wave plate (QWP), half wave plate (HWP) and a polarizing beam splitter (PBS) fulfill the necessary state projections.}
\label{fig:Polmeasure}
\end{figure}

To create polarization-OAM entanglement of two indistinguishable photons travelling in the same direction (cf. Eq.~(\ref{eq:dualPolOAM1})) a two-photon source has to be employed, e.g., using parametric down conversion. The produced down converted polarization-path entangled photon pairs are then described by the quantum state given in Eq.~(\ref{eq:dualHVpath1}). Since in our case we want to replace the spatial degree of freedom by OAM variables, two different polarization to OAM transferrers ($\pi\rightarrow l$) are applied thereafter, where $\pi$ ($l$) denotes the polarization (OAM quantum number) of the particle, followed by a PBS for spatial mode mixing.\\
A possible $\pi\rightarrow l$ transferrer described in \cite{Nagali(2009)} consists of a quarter wave plate (QWP) changing $H$ ($V$) polarization to right (left) circular polarization, a q-plate and a PBS. A q-plate transforms left (right) circular polarization to right (left) circular polarization while at the same time producing OAM of $l$ ($-l$) \cite{Marrucci(2006)}. Recently, q-plates producing OAM with values of up to $|l|=100$ have been reported \cite{DAmbrosio(2013)}. Measuring only the horizontal output mode of the q-plate by use of the PBS, every polarization state $1/\sqrt{2}(\ket{H}+\ket{V})$ with vanishing OAM ($\ket{l=0}$) will be transformed into the following superposition state of OAM $\pm l$ \cite{Nagali(2009)}
\begin{equation}
	\frac{1}{\sqrt{2}}\left(\ket{H}+\ket{V}\right)\ket{l=0} \ \stackrel{\text{q-plate}}{\longrightarrow} \ \frac{1}{\sqrt{2}}\left(\ket{l}+\ket{-l}\right)\ket{H} \ .
\label{eq:transferrer1}
\end{equation}
In principle the same result can be achieved by using a second type of $\pi\rightarrow l$ transferrer as described in \cite{Fickler(2012)}. Hereby, spatial light modulators (SLM) are employed which can produce OAM with values of up to $|l|=300$. By separating the photons according to their polarization and sending them on different SLM, OAM of $l$ ($-l$) can be transferred independently to $H$ ($V$) polarized photons. The subsequent recombination of the distinct photon modes and deleting the polarization degree of freedom by use of a linear polarizer of diagonal polarization $D_{+}$ ($\ket{D_{+}}=1/\sqrt{2}(\ket{H}+\ket{V}$)) completes the entanglement transfer \cite{Fickler(2012)}
\begin{equation}
\begin{aligned}
	\frac{1}{\sqrt{2}}\left(\ket{H}+\ket{V}\right)\ket{l=0} \ & \stackrel{\text{SLM}}{\longrightarrow} \ \frac{1}{\sqrt{2}}\left(\ket{H,l}+\ket{V,-l}\right) \\
	& \stackrel{D_{+}}{\longrightarrow} \ \frac{1}{\sqrt{2}}\left(\ket{l}+\ket{-l}\right)\ket{D_{+}} \ .
\label{eq:transferrer2} \\
\end{aligned}
\end{equation}
The output states of the two $\pi\rightarrow l$ transferrers only differ in their polarization (cf. Eq.~(\ref{eq:transferrer1}) and (\ref{eq:transferrer2})). Half wave plates (HWP) of angle $\theta_{\text{HWP}1}=22.5^{\circ}$ can then be used to rotate the polarization $D_{+}$ to $H$. Another possibility would be to rotate the coordinate system around an angle of $\theta_{sys}=45^{\circ}$; however, for a rotated coordinate system all subsequent devices have to be rotated around the same angle $\theta_{sys}$ as well.\\
In both cases of $\pi\rightarrow l$ transferrers the polarization has to be changed in one of the two spatial modes (e.g., $\vec{k}_{2}$). This is done by employing a HWP of angle $\theta_{\text{HWP}2}=45^{\circ}$, which finally produces a polarization-OAM entangled state but still in two spatial path modes (see Fig.~\ref{fig:source2})
\begin{equation}
\begin{aligned}
	\ket{\Psi'}= \frac{1}{\sqrt{2}}\left(\ket{H,l,\vec{k}_{1}} \ket{V,-l,\vec{k}_{2}}+\ket{H,-l,\vec{k}_{1}}\ket{V,l,\vec{k}_{2}}\right) \ .
\label{eq:statePolOAMPath}
\end{aligned}
\end{equation}
\\
Since the state given in Eq.~(\ref{eq:statePolOAMPath}) displays two-photon entanglement for either variables, OAM and path, or polarization and OAM, several tests of entanglement are possible. For example, OAM-path entanglement could be tested in an analogous way as the path-polarization experiments presented in \cite{Bose(2013)}. However, due to the simultaneous entanglement of OAM, polarization and path, the two photons could not interfere when using polarization sensitive photon-detectors since in this case the photons become distinguishable. On the other hand, projecting all photons onto diagonal polarized states would project the complete state onto a maximally OAM-path entangled state. In this case interference becomes possible  even for polarization sensitive detectors. Another experiment could be to test polarization-OAM entanglement of the two photons still being in distinct spatial modes. This measurement would require entanglement sorting via polarization and OAM, as being suggested in the following for two photons travelling along the same direction.\\
In what follows we thus concentrate on the two-photon state prepared in only one single path. To this aim the additional spatial degree of freedom has to be removed. Since photons in mode $\vec{k}_{1}$ are always horizontally polarized and photons in mode $\vec{k}_{2}$ are always vertically polarized, all photons can be projected onto a single path mode by mixing modes $\vec{k}_{1}$ and $\vec{k}_{2}$ via a PBS (see Fig.~\ref{fig:source2}). This eliminates the spatial degree of freedom and creates entanglement solely between polarization and OAM variables of the form (cf. Eq.~(\ref{eq:dualPolOAM1}))
\begin{equation}
\begin{aligned}
	\ket{\Psi} & = \frac{1}{\sqrt{2}}\left(\ket{H,l}\right. \left. \ket{V,-l}+\ket{V,l}\ket{H,-l}\right) \\
	& \equiv \frac{1}{\sqrt{2}}\left(\ket{l,H} \ket{-l,V}+\ket{-l,H}\ket{l,V}\right) \ .
\label{eq:statePolOAM}
\end{aligned}
\end{equation}
Starting from Eq.~(\ref{eq:statePolOAM}) we next make use of the properties of entanglement duality \cite{Bose(2013)}. As it is shown in Eq.~(\ref{eq:statePolOAM}) the polarization-OAM entanglement can be written in two equivalent ways, by interchanging the variables ($H,V$) and ($l,-l$). This has been interpreted as two distinct possibilities of labeling the entanglement of two identical particles \cite{Bose(2013)}. Both entanglement labelings display two-particle entanglement. Now, the produced dual state can be separated into two spatial modes depending on its labeling variables, i.e., either via the polarization or the OAM degrees of freedom. This so-called entanglement sorting allows for a direct measurement of the corresponding variables' entanglement (see Fig.~\ref{fig:OAMmeasure} and \ref{fig:Polmeasure}) as in both cases experimentally implementable witnesses can be set up to prove the entanglement. Hereby, an observable $\hat{W}$ is called entanglement witness when indicating entanglement in the following way \cite{Guhne(2009)}
\begin{equation}
\begin{aligned}
\text{Tr}[\hat{W}\rho_{s}] \geq 0 \ , \ \ \ \ \
\text{Tr}[\hat{W}\rho_{e}] < 0 \ ,
\end{aligned}
\end{equation}
with $\rho_{s}$ representing all separable states and $\rho_{e}$ representing at least one entangled state.

\section{Entanglement Sorting via Polarization}
\label{sec:3}

\begin{figure}[b]
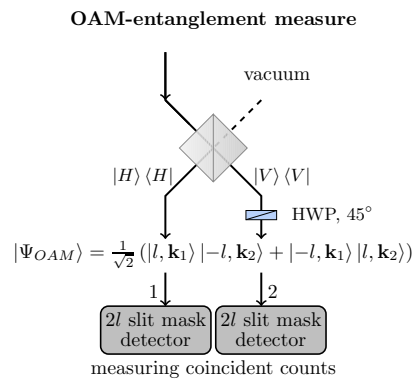

\centering
\scalebox{0.75}{
\include{OAMmeasure_new}}
\caption{Separating the polarization-OAM entangled state (cf. Eq.~(\ref{eq:statePolOAM})) due to its polarization variables via PBS and eliminating the polarization degree of freedom by use of a HWP leads to the OAM-entangled state $\ket{\Psi_{OAM}}$. A given entanglement witness $\hat{W}$ (cf. Eq.~(\ref{eq:WitnessOAM})) can be determined measuring coincidence counts of different state projections whereby state projections are accomplished by radially symmetric slit masks.}
\label{fig:OAMmeasure}
\end{figure}

Sorting the state via polarization by use of a PBS divides the given state $\ket{\Psi} $ of Eq.~(\ref{eq:statePolOAM}) into $H$ and $V$ polarized photons (see Fig.~\ref{fig:OAMmeasure}). Then flipping polarization in the vertical mode from $V$ to $H$ employing a HWP of $\theta=45^{\circ}$ eliminates the polarization degree of freedom and produces the OAM-path entangled state $\ket{\Psi_{OAM}}$ of Eq.~(\ref{eq:dualOAMpath1}). This allows to measure OAM-path entanglement. 
As shown in \cite{Fickler(2012)} a subsequent combination of a radial symmetric slit mask ($2l$ slits) and a bucket detector in each spatial output mode $1$ and $2$ allows for measuring the OAM-entanglement using an appropriate entanglement witness $\hat{W}$ and considering coincident counts only, even for very high quantum numbers $l$. In the following the experimental idea from \cite{Fickler(2012)} will be explained, but a different entanglement witness will be employed. The derivation of this entanglement witness can be found in \cite{Guhne(2002),Guhne(2009)}.
\\
For the experimental approach one can make use of the fact that any state of the form \cite{Fickler(2012)}
\begin{equation}
\ket{\Psi_{j}}= \frac{1}{\sqrt{2}}\ (\ket{l,\vec{k}_{j}}+e^{i\varphi_{j}}\ket{-l,\vec{k}_{j}}) \ ,
\label{eq:State2lInt}
\end{equation}
displays a radial intensity profile with $2l$ intensity maxima arranged in a circle, where $j=1,2$ denotes the spatial path mode. Therefore, a rotatable slit mask of the same symmetry can be used to measure every possible superposition as a function of the mask's rotation angle $\gamma'_{j}$
. Obviously, the state measured by the mask then reads \cite{Fickler(2012)}
\begin{equation}
\begin{aligned}
\ket{\chi_{j}(\phi_{j})}=\frac{1}{\sqrt{2}}\ (\ket{l,\vec{k}_{j}}+e^{i\phi_{j}}\ket{-l,\vec{k}_{j}}) \ ,
\label{eq:statemasktext}
\end{aligned}
\end{equation}
where the phase factor $\phi_{j}$ of the projected state is determined by the angular position of the slit mask \cite{Fickler(2012)}
\begin{equation}
	\gamma'_{j}=\frac{\phi_{j}\cdot 360^{\circ}}{2l\cdot 2\pi} \Leftrightarrow  \phi_{j}=\frac{\gamma'_{j} \cdot 2l \cdot 2\pi}{360^{\circ}} \ . 
\end{equation}
Coincidence measurements of both modes can then be performed as a function of the angular positions $\gamma'_{1}$ and $\gamma'_{2}$ of the slit masks in the two output modes.\\
The entanglement witness can be formulated in an experimentally implementable way consisting of different mutually unbiased measures \cite{Guhne(2002)}
\begin{equation}
\begin{aligned}
\hat{W}= & \frac{1}{2}\mathbb{1} -\ket{\Psi_{OAM}}\bra{\Psi_{OAM}} \\
	= & \frac{1}{2} \left( \ket{\,l_{1},l_{2}}\bra{l_{1},l_{2}\,} + \ket{-l_{1},-l_{2}}\bra{-l_{1},-l_{2}} \right. \\
	 & - \ket{{d_{+}}_{1},{d_{+}}_{2}}\bra{{d_{+}}_{1},{d_{+}}_{2}} - \ket{{d_{-}}_{1},{d_{-}}_{2}}\bra{{d_{-}}_{1},{d_{-}}_{2}} \\
	 & \left. + \ket{\mathcal{L}_{1},\mathcal{R}_{2}}\bra{\mathcal{L}_{1},\mathcal{R}_{2}} + \ket{\mathcal{R}_{1},\mathcal{L}_{2}}\bra{\mathcal{R}_{1},\mathcal{L}_{2}} \right) \ ,
	\label{eq:Witness}
\end{aligned}
\end{equation}
where $\ket{A_{1},B_{2}}\equiv \ket{A,\vec{k}_{1}}\ket{B,\vec{k}_{2}}$. All parts of the witness can be expressed by projected states of the slit masks (cf. Eq.~(\ref{eq:statemasktext})). This can be accomplished by simply adjusting the slit masks to appropriate angular positions. From this follow the identities $\ket{{d_{+}}_{j}}=\ket{\chi_{j}(0)}$, $\ket{{d_{-}}_{j}}=\ket{\chi_{j}(\pi)}$ denoting diagonal OAM states, and $\ket{\mathcal{L}_{j}}=\ket{\chi_{j}(\pi/2)}$ and $\ket{\mathcal{R}_{j}}=\ket{\chi_{j}(-\pi/2)}$ denoting circular OAM states in analogy to polarization states. The complete witness in terms of the mask projections now reads (cf. Eq.~(\ref{eq:Witness}))
\begin{equation}
\begin{aligned}
\hat{W}= & \frac{1}{2}\mathbb{1} -\ket{\Psi_{OAM}}\bra{\Psi_{OAM}} \\
	= & \frac{1}{2} \big(\ \ket{\,l_{1},l_{2}}\bra{l_{1},l_{2}\,} + \ket{-l_{1},-l_{2}}\bra{-l_{1},-l_{2}}  \\
	 &\hspace{4mm} \ - \ket{\chi_{1}(0),\chi_{2}(0)}\bra{\chi_{1}(0),\chi_{2}(0)} \\
	 &\hspace{4mm} \ - \ket{\chi_{1}(\pi),\chi_{2}(\pi)}\bra{\chi_{1}(\pi),\chi_{2}(\pi)} \\
	 &\hspace{4mm} \ + \ket{\chi_{1}(\pi/2),\chi_{2}(-\pi/2)}\bra{\chi_{1}(\pi/2),\chi_{2}(-\pi/2)} \\
	 &\hspace{4mm} \  + \ket{\chi_{1}(-\pi/2),\chi_{2}(\pi/2)}\bra{\chi_{1}(-\pi/2),\chi_{2}(\pi/2)}\ \big) \ .
	\label{eq:WitnessProjections}
\end{aligned}
\end{equation}
Since in the experiment one is measuring intensities, i.e., coincidence counts, the measurements have to be normalized by the sum of the intensities of the basis states $\ket{l_{1},l_{2}}$, $\ket{l_{1},-l_{2}}$, $\ket{-l_{1},l_{2}}$ and $\ket{-l_{1},-l_{2}}$ \cite{Barbieri(2003)}.\\
Calculating the theoretical expectation value of the entanglement witness (Eq.~(\ref{eq:Witness})) and the maximally entangled state (Eq.~(\ref{eq:dualOAMpath1})) gives a value of
\begin{equation}
	\bra{\Psi_{OAM}}\hat{W}\ket{\Psi_{OAM}} = - \frac{1}{2} \ .
\end{equation}
Similar to \cite{Fickler(2012)} we now want to show that a general separable state (cf. Eq.~(\ref{eq:State2lInt}))
\begin{equation}
\begin{aligned}
	\ket{\Psi_{s}}=&\ket{\Psi_{1}',\Psi_{2}'} \\
	 =&\left(a\ket{l,\vec{k}_{1}}\vphantom{e^{i\varphi_{1}}}+b e^{i\varphi_{1}}\ket{-l,\vec{k}_{1}}\right) \\ 
	& \hspace{15mm} \otimes\left(c\ket{l,\vec{k}_{2}} + d e^{i\varphi_{2}}\ket{-l,\vec{k}_{2}}\right) ,
\label{eq:statePsi12}
\end{aligned}
\end{equation}
does \textit{not} violate the entanglement witness, where $a,b,c,d\in \mathbb{R}$ and $a^{2}+b^{2}=c^{2}+d^{2}=1$. For this to prove we have to determine the expectation value of the witness $\bra{\Psi_{s}}\hat{W}\ket{\Psi_{s}}$ and show that $\bra{\Psi_{s}}\hat{W}\ket{\Psi_{s}} \geq 0$.
For the state $\ket{\Psi_{s}}$ from Eq.~(\ref{eq:statePsi12}) the components of the entanglement witness (cf. Eq.~(\ref{eq:Witness})) take the following form
\begin{equation}
\begin{aligned}
	 \braket{\Psi_{s}|\;l_{1},l_{2}\,|\Psi_{s}} &= a^{2}c^{2} \ , \\
	 \braket{\Psi_{s}|-l_{1},-l_{2}|\Psi_{s}} &= b^{2}d^{2} \ , \\
	 \braket{\Psi_{s}|{d_{+}}_{1},{d_{+}}_{2}|\Psi_{s}} &= | \frac{1}{2} (a+be^{i\varphi_{1}})(c+de^{i\varphi_{2}}) |^{2} \ , \\
	 \braket{\Psi_{s}|{d_{-}}_{1},{d_{-}}_{2}|\Psi_{s}} &= | \frac{1}{2} (a-be^{i\varphi_{1}})(c-de^{i\varphi_{2}}) |^{2} \ , \\
	 \braket{\Psi_{s}|\mathcal{L}_{1},\mathcal{R}_{2}|\Psi_{s}} &= | \frac{1}{2} (a-ibe^{i\varphi_{1}})(c+ide^{i\varphi_{2}}) |^{2} \ , \\
	 \braket{\Psi_{s}|\mathcal{R}_{1},\mathcal{L}_{2}|\Psi_{s}} &= | \frac{1}{2} (a+ibe^{i\varphi_{1}})(c-ide^{i\varphi_{2}}) |^{2} \ ,
\label{eq:WitnessOAM}
\end{aligned}
\end{equation}
where $\braket{\Psi_{s}|A_{1},B_{2}|\Psi_{s}} = \braket{\Psi_{s}|A_{1},B_{2}}\braket{A_{1},B_{2}|\Psi_{s}} $. Solving the absolute squares
\begin{equation}
\begin{aligned}
	 \braket{\Psi_{s}|{d_{+}}_{1},{d_{+}}_{2}|\Psi_{s}} = &  \frac{1}{4} [1 + 2ab \cos(\varphi_{1}) + 2cd \cos(\varphi_{2}) \\
  & \hspace{14mm}  + 4abcd \cos(\varphi_{1}) \cos(\varphi_{2})] \ , \\
	 \braket{\Psi_{s}|{d_{-}}_{1},{d_{-}}_{2}|\Psi_{s}} = &  \frac{1}{4} [1 - 2ab \cos(\varphi_{1}) - 2cd \cos(\varphi_{2}) \\
	& \hspace{14mm}  + 4abcd \cos(\varphi_{1}) \cos(\varphi_{2})] \ , \\
	 \braket{\Psi_{s}|\mathcal{L}_{1},\mathcal{R}_{2}|\Psi_{s}} = &  \frac{1}{4} [1 + 2ab \sin(\varphi_{1}) - 2cd \sin(\varphi_{2}) \\
  & \hspace{14mm}  - 4abcd \sin(\varphi_{1}) \sin(\varphi_{2})] \ , \\
	 \braket{\Psi_{s}|\mathcal{R}_{1},\mathcal{L}_{2}|\Psi_{s}} = &  \frac{1}{4} [1 - 2ab \sin(\varphi_{1}) + 2cd \sin(\varphi_{2}) \\
	& \hspace{14mm}  - 4abcd \sin(\varphi_{1}) \sin(\varphi_{2})] \ , \\
\end{aligned}
\end{equation}
leads to the following value for the entanglement witness (cf. Eq.~(\ref{eq:Witness}))
\begin{equation}
\begin{aligned}
\bra{\Psi_{s}}\hat{W}\ket{\Psi_{s}} = & \; \frac{1}{2} \; [ a^{2}c^{2} + b^{2}d^{2} - 2abcd\cos(\varphi_{1})\cos(\varphi_{2}) \\ 
	 & \hspace{24mm} - 2abcd\sin(\varphi_{1})\sin(\varphi_{2}) ] \ .
	\label{eq:WitnessSep}
\end{aligned}
\end{equation}
Eq.~(\ref{eq:WitnessSep}) reaches a minimal value of $0$ (for $\varphi_{1}=\varphi_{2}$, $a=d$ and $b=c$) what confirms the validity of the witness and the OAM-path entanglement of the two-photon state $\ket{\Psi_{OAM}}$.\\
Note that for very high quantum numbers $l$ the coincidence signal might have to be corrected by subtracting accidental coincident counts. However, experiments for at least $l=100$ still display very good results without correction \cite{Fickler(2012)}.\\
Another possibility for measuring the OAM-entanglement witness has been implemented consisting of SLM in the two spatial output modes \cite{Agnew(2012)}. In this case SLM take the role of the slit masks and project the OAM-modes onto the distinct superposition states what allows for measuring the witness.

\section{Entanglement sorting via Angular Momentum}
\label{sec:4}
The second possibility of entanglement sorting requires at first to divide up the polarization-OAM entangled state of Eq.~(\ref{eq:statePolOAM}) according to the OAM numbers $\pm l$. Holographic fork masks of appropriate order $l$ can be used to achieve this separation \cite{Heckenberg(1992)}, as they diffract incident photons while at the same time change their OAM. In the first diffraction order photons with OAM-change of $\Delta l=+ l$ can be found in one direction and photons with OAM-change $\Delta l= -l$ in the other direction. Coupling the photons into single mode fibers, transmitting photons with $l=0$ only, allows for separation and detection of photons with OAM $\pm l$ \cite{Nagali(2009)}.
The state after the separation due to OAM components carries the polarization entanglement and corresponds to $\ket{\Psi_{Pol}}$ of Eq.~(\ref{eq:dualHVpath1}).
Proving polarization-entanglement can now be accomplished by using again an appropriate entanglement witness. In \cite{Barbieri(2003)} this approach has been used to verify entanglement for a polarization singlet state. Here the witness for testing the given state (cf. Eq.~(\ref{eq:dualHVpath1})) is of the same form as Eq.~(\ref{eq:Witness}) but formulated in terms of polarization \cite{Guhne(2002)}
\begin{equation}
\begin{aligned}
\hat{W}= & \frac{1}{2}\mathbb{1} -\ket{\Psi_{Pol}}\bra{\Psi_{Pol}} \\
	= & \frac{1}{2} \left( \ket{H_{1},H_{2}}\bra{H_{1},H_{2}} + \ket{V_{1},V_{2}}\bra{V_{1},V_{2}} \right. \\
	 & -\! \ket{{D_{+}}_{1},{D_{+}}_{2}}\bra{{D_{+}}_{1},{D_{+}}_{2}}\! -\! \ket{{D_{-}}_{1},{D_{-}}_{2}}\bra{{D_{-}}_{1},{D_{-}}_{2}} \\
	 & \left. + \ket{L_{1},R_{2}}\bra{L_{1},R_{2}} + \ket{R_{1},L_{2}}\bra{R_{1},L_{2}} \right) \ ,
	\label{eq:WitnessPol}
\end{aligned}
\end{equation}
where again $\ket{A_{1},B_{2}}\equiv \ket{A,\vec{k}_{1}}\ket{B,\vec{k}_{2}}$ and the diagonal polarization states $\ket{D_{\pm}}=(\ket{H}\pm\ket{V})/\sqrt{2}$ and the circular polarization states left $\ket{L}=(\ket{H}+i\ket{V})/\sqrt{2}$ and right $\ket{R}=(\ket{H}-i\ket{V})/\sqrt{2}$ have been inserted.\\
The different combinations of polarizations in the two distinct output modes can be measured by applying QWP, HWP and a PBS sequentially \cite{Barbieri(2003)} where again coincident counts have to be taken into account. Additionally, the measurements have to be normalized by the sum of the coincident rates in the basis $\ket{H_{1},H_{2}}$, $\ket{H_{1},V_{2}}$, $\ket{V_{1},H_{2}}$ and $\ket{V_{1},V_{2}}$ in order to obtain probabilities \cite{Barbieri(2003)}.

\section{Generalization to three- and four-photon entanglement}
\label{sec:5}

The combination of OAM and polarization degree of freedom also allows for implementing entanglement sorting with more than two identical photons. In \cite{Bouwmeester(1999)} it was shown that three-photon polarization-entanglement can be produced in the form of a Greenberger-Horne-Zeilinger (GHZ) state
\begin{equation}
\begin{aligned}
	&\ket{\Psi_{GHZ}}=\frac{1}{\sqrt{2}}\left(\ket{H,\vec{k}_{1}}\ket{V,\vec{k}_{2}}\ket{V,\vec{k}_{3}} \right. \\ 
	& \hspace{30mm} \left. +\ket{V,\vec{k}_{1}}\ket{H,\vec{k}_{2}}\ket{H,\vec{k}_{3}}\right) \ ,
\label{eq:GHZ}
\end{aligned}
\end{equation}
where the photons are propagating in three different spatial modes. Employing $\pi\rightarrow l$ transferrers (cf. Eq.~(\ref{eq:transferrer1}) or (\ref{eq:transferrer2})) and a polarization flipper leads to a three-photon polarization-OAM-path entanglement. All photons can now be combined into one single path by use of a collective lens, which gives rise to the following state
\begin{equation}
\begin{aligned}
	&\ket{\Psi_{GHZ}'}\\ 
	&\hspace{5mm}=\frac{1}{\sqrt{2}}\left(\ket{H,l}\ket{V,-l}\ket{V,-l}+\ket{V,l}\ket{V,l}\ket{H,-l}\right) \\
	&\hspace{5mm}\equiv\frac{1}{\sqrt{2}}\left(\ket{l,H}\ket{-l,V}\ket{-l,V}+ \ket{-l,H}\ket{l,V}\ket{l,V}\right)  .
\label{eq:threephotonstate1}
\end{aligned}
\end{equation}
Here, entanglement sorting via polarization produces the OAM-path entangled state
\begin{equation}
\begin{aligned}
	&\ket{\Psi_{OAM,3}}=\frac{1}{\sqrt{2}} \left( \ket{l,\vec{k}_{1}}\ket{-l,\vec{k}_{2}}\ket{-l,\vec{k}_{2}} \right. \\ 
	& \hspace{30mm} + \left. \ket{-l,\vec{k}_{1}}\ket{l,\vec{k}_{2}}\ket{l,\vec{k}_{2}} \right) \ ,
\end{aligned}
\end{equation}
whereas entanglement sorting via OAM produces polarization-path entanglement of the form
\begin{equation}
\begin{aligned}
	&\ket{\Psi_{Pol,3}}=\frac{1}{\sqrt{2}} \left( \ket{H,\vec{k}_{1}}\ket{V,\vec{k}_{2}}\ket{V,\vec{k}_{2}} \right. \\
	& \hspace{30mm} + \left. \ket{V,\vec{k}_{1}}\ket{V,\vec{k}_{1}}\ket{H,\vec{k}_{2}} \right) \ .
\end{aligned}
\end{equation}
Obviously, entanglement sorting for three-photon entanglement can only be accomplished in a modified way since the asymmetric photon distribution causes the sorted states to be of asymmetric form.\\
In case of four-photon entanglement which could arise from type-II parametric down-conversion four photons are created in two distinct directions \cite{Weinfurter(2001),Weinfurter(2003)}
\begin{equation}
\begin{aligned}
	\ket{\Psi_{Pol,4}} =  \frac{1}{\sqrt{3}} \left( \vphantom{\ket{H,\vec{k}_{1}}} \right.\!\! & \left.  \ket{H,\vec{k}_{1}}\ket{H,\vec{k}_{1}}\ket{V,\vec{k}_{2}}\ket{V,\vec{k}_{2}} \right.  \\
	  + & \left. \ket{V,\vec{k}_{1}}\ket{V,\vec{k}_{1}}\ket{H,\vec{k}_{2}}\ket{H,\vec{k}_{2}} \right. \\
	 \vphantom{\frac{1}{\sqrt{3}} } + & \left. \ket{H,\vec{k}_{1}}\ket{V,\vec{k}_{1}}\ket{H,\vec{k}_{2}}\ket{V,\vec{k}_{2}} \right) \ .
\label{eq:Polfour}
\end{aligned}
\end{equation}
By using SLM to induce OAM of $l$ and $-l$ in spatial modes $\vec{k}_{1}$ and $\vec{k}_{2}$, respectively, and collecting all photons in a single path mode with an optical lens, the following  four-photon polarization-OAM entangled state is generated
\begin{equation}
\begin{aligned}
	\ket{\Psi_{4}} = \frac{1}{\sqrt{3}} \left( \vphantom{\ket{H,l}} \right. \!\! & \left. \ket{H,l}\ket{H,l}\ket{V,-l}\ket{V,-l} \right.  \\
	 + & \left. \ket{V,l}\ket{V,l}\ket{H,-l}\ket{H,-l} \right. \\
	 \vphantom{\frac{1}{\sqrt{3}} } + & \left.\ket{H,l}\ket{V,l}\ket{H,-l}\ket{V,-l} \right) \ .
\label{eq:PolOAMfour}
\end{aligned}
\end{equation}
Due to the symmetry of the four-photon state dual entanglement sorting can be accomplished in the same way as in the case of two-photon polarization-OAM entanglement. For example, applying entanglement sorting via polarization by use of a PBS leads to OAM-path entanglement (cf. Eq.~(\ref{eq:Polfour}))
\begin{equation}
\begin{aligned}
	\ket{\Psi_{OAM,4}} =  \frac{1}{\sqrt{3}} \left( \vphantom{\ket{l,\vec{k}_{1}}} \right. \!\! & \left. \ket{l,\vec{k}_{1}}\ket{l,\vec{k}_{1}}\ket{-l,\vec{k}_{2}}\ket{-l,\vec{k}_{2}} \right. \\
	  + & \left. \ket{-l,\vec{k}_{1}}\ket{-l,\vec{k}_{1}}\ket{l,\vec{k}_{2}}\ket{l,\vec{k}_{2}} \right. \\
	  \vphantom{\frac{1}{\sqrt{3}} } + & \left. \ket{l,\vec{k}_{1}}\ket{-l,\vec{k}_{1}}\ket{l,\vec{k}_{2}}\ket{-l,\vec{k}_{2}} \right) \ ,
\label{eq:OAMfour}
\end{aligned}
\end{equation}
whereas entanglement sorting via OAM by use of a fork hologram projects the state $\ket{\Psi_{4}}$ (cf. Eq.~(\ref{eq:PolOAMfour})) onto the polarization-path entangled state $\ket{\Psi_{Pol,4}}$ (cf. Eq.~(\ref{eq:Polfour})).

\section{Conclusion}
\label{sec:6}
In conclusion we presented in Sec.~\ref{sec:2} an experimental setup able to produce polarization-OAM entangled two-photon states where both photons propagate in the same direction. For these states we demonstrated in sections~\ref{sec:3} and ~\ref{sec:4} that due to the duality of the identical photons entanglement sorting can be accomplished depending on their angular momenta, i.e., via their polarization or OAM degrees of freedom.\\
In Sec.~\ref{sec:3} we showed that entanglement sorting via polarization can be achieved using a PBS while in Sec.~\ref{sec:4} a holographic fork mask was employed to achieve entanglement sorting via OAM. Hence, we could show that simple experimental setups can be used to perform entanglement sorting and, thereby, prove the dual entanglement of the two angular momenta degrees of freedom.\\
In Sec.~\ref{sec:5} we finally pointed out that entanglement sorting can be generalized to higher photon numbers, discussing in particular the cases $N=3$ and $N=4$. From these discussions it became evident that in case of even photon numbers entanglement duality occurs and symmetric entanglement sorting is possible -- meaning that the two entanglement sorted states have symmetric path-entangled structure -- whereas in case of odd photon numbers the entanglement sorted states are of asymmetric form.

\bibliography{literatur2014oct}
\bibliographystyle{apsrev4-1}

\end{document}

%% file: source2small.tex
\begin{tikzpicture}[scale=1.7]
	\tikzstyle{every node}=[font=\normalsize]

\filledboxtwo{-0.35}{-0.2+0.7}{0.35}{0.2+0.7};
\node[below] at (0.,0.11+0.7){source};

\node[below] at (0,+0.03){$\ket{\Psi_{Pol}}=\frac{1}{\sqrt{2}}\left(\ket{H,\vec{k}_{1}}\ket{V,\vec{k}_{2}}+\ket{V,\vec{k}_{1}}\ket{H,\vec{k}_{2}}\right)$};

\filledboxtwo{-0.4-0.5}{-0.2-0.7}{0.4-0.5}{0.2-0.7};
\node[below] at (-0.5,0.12-0.7){$\pi\rightarrow l$};
\filledboxtwo{-0.4+0.5}{-0.2-0.7}{0.4+0.5}{0.2-0.7};
\node[below] at (0.5,0.12-0.7){$\pi\rightarrow l$};

\draw[->,line width=1pt] (-0.2,-0.2+0.7) to ++(-0.3,-0.3) to ++(0,-0.15);
\node[left] at (-0.35,-0.3+0.7){$1$};
\draw[->,line width=1pt] (0.2,-0.2+0.7) to ++(0.3,-0.3) to ++(0,-0.15);
\node[right] at (0.35,-0.3+0.7){$2$};

\draw[->,line width=1pt] (-0.5,-0.35) to ++(-0.0,-0.15);
\draw[->,line width=1pt] (0.5,-0.35) to ++(0.0,-0.15);

\draw[line width=1pt,->] (-0.5,-0.5-0.4) to ++(0,-0.5);
\draw[line width=1pt,->] (0.5,-0.5-0.4) to ++(0,-0.5);

\draw [rotate around={0:(0.5,-1.6)},fill=blue1] (0.5-0.16,-1.6-0.048+0.5) rectangle ++(0.32,0.096);
\draw [rotate around={0:(0.5,-1.6)}] (0.5-0.16,-1.6-0.048+0.5)--++(0.32,0.096);
\node[right] at(1.3-0.5,-1.6+0.5) {HWP, $45^{\circ}$};

\node[below] at (0,-1.4-0.05){$\ket{\Psi'}= \frac{1}{\sqrt{2}}\left(\ket{H,l,\vec{k}_{1}} \ket{V,-l,\vec{k}_{2}}+\ket{H,-l,\vec{k}_{1}}\ket{V,l,\vec{k}_{2}}\right)$};

\draw[line width=1pt,->] (-0.5,-1.4-0.2-0.3) to ++(0,-0.2);
\draw[line width=1pt,-] (-0.5,-1.4-0.2-0.3) to ++(0,-0.2) to ++(0.5,-0.5);
\draw[line width=1pt,->] (0.5,-1.4-0.2-0.3) to ++(0,-0.2);
\draw[line width=1pt,-] (0.5,-1.4-0.2-0.3) to ++(0,-0.2) to ++(-1,-1);

\draw[line width=1pt,dashed] (-0.5+0.5,-1.4-0.2-0.3-0.2-0.5) to ++(0.5,-0.5);
\node[right] at(-0.5+0.5+0.5,-1.4-0.2-0.3-0.2-0.5-0.5) {vacuum};
\draw[line width=1pt,->] (0.5,-1.4-0.2-0.3-0.2) to ++(-1,-1)to ++(0,-0.2);

\pbs{0}{-1.4-0.2-0.3-0.2-0.5}{45};
\node[right] at(0.4,-1.4-0.2-0.3-0.2-0.5) {PBS};

\draw[line width=1pt,->] (-0.5,-1.4-0.2-0.3-0.2-1) to ++(0,-0.2);

\node[below] at (0,-1.4-0.2-0.3-0.2-1-0.2-0.05){$\ket{\Psi}=\frac{1}{\sqrt{2}}\left(\ket{H,l}\ket{V,-l}+\ket{H,-l}\ket{V,l}\right)$};

\end{tikzpicture}

%% file: OAMmeasure_new.tex
\begin{tikzpicture}[scale=1.7]
	\tikzstyle{every node}=[font=\normalsize]

\node[below] at (0,0.5){\bf OAM-entanglement measure};

\draw[line width=1pt,->] (-0.5,0) to ++(0,-0.5);
\draw[line width=1pt,-] (-0.5,0) to ++(0,-0.5) to ++(0.25,-0.25);

\draw[line width=1pt,->] (-0.5,-0.5) to ++(1,-1)to ++(0,-1+0.5+0.35-0.1-0.2);
\node[right] at (0.25,-0.25) {vacuum};
\draw[line width=1pt, ->] (0,-1) to ++(-0.5,-0.5)to ++(0,-1+0.5+0.35-0.1-0.2);
\draw[line width=1pt,dashed] (0.5,-0.5) to ++(-0.5,-0.5);

\pbs{0}{-1.}{45};

\node[right] at (0.35,-1.3) {\small $\ket{V}\bra{V}$};
\node[left] at (-0.35,-1.3) {\small $\ket{H}\bra{H}$};

\begin{scope}[yshift=-6.0mm]
\draw [rotate around={0:(0.5,-1.6)},fill=blue1] (0.5-0.16,-1.6-0.048+0.5) rectangle ++(0.32,0.096);
\draw [rotate around={0:(0.5,-1.6)}] (0.5-0.16,-1.6-0.048+0.5)--++(0.32,0.096);
\node[right] at(1.3-0.55,-1.6+0.5) {\small HWP, $45^{\circ}$};

\end{scope}

\node[below] at (0,-1.1-1+0.5-0.1-0.2){$\ket{\Psi_{OAM}}=\frac{1}{\sqrt{2}}\left(\ket{l,\vec{k}_{1}}\ket{-l,\vec{k}_{2}}+\ket{-l,\vec{k}_{1}}\ket{l,\vec{k}_{2}}\right)$};

\draw[line width=1pt,->] (-0.5,-1.1-1+0.5-0.4-0.1-0.2) to ++(0,-1+0.5+0.35-0.2+0.1-0.1);
\node[left] at(-0.5,-1.1-1+0.5-0.4-0.1-1+0.5+0.35-0.15+0.1-0.2) {1};
\draw[line width=1pt,->] (0.5,-1.1-1+0.5-0.4-0.1-0.2) to ++(0,-1+0.5+0.35-0.2+0.1-0.1);
\node[right] at(0.5,-1.1-1+0.5-0.4-0.1-1+0.5+0.35-0.15+0.1-0.2) {2};

\filledboxtwo{0+0.1-0.075}{-4+0.5+1.5-0.35-0.1-0.2}{1+0.1+0.075}{-4.5+0.5+1.5-0.35-0.1-0.2};
\node[below] at (0.5+0.1,-4+0.5+1.5-0.35-0.1-0.2){$2l$ slit mask};
\node[below] at (0.5+0.1,-4.2+0.5+1.5-0.35-0.1-0.2){detector};
\filledboxtwo{-1-0.1-0.075}{-4+0.5+1.5-0.35-0.1-0.2}{0-0.1+0.075}{-4.5+0.5+1.5-0.35-0.1-0.2};
\node[below] at (-0.5-0.1,-4+0.5+1.5-0.35-0.1-0.2){$2l$ slit mask};
\node[below] at (-0.5-0.1,-4.2+0.5+1.5-0.35-0.1-0.2){detector};

\node[below] at (0,-4.5+0.5+1.5-0.35-0.1-0.2){measuring coincident counts};

\end{tikzpicture}

%% file: Polmeasure_new.tex
\begin{tikzpicture}[scale=1.7]
	\tikzstyle{every node}=[font=\normalsize]


\begin{scope}[xshift=0mm]

\node[below] at (0,0.5){\bf polarization-entanglement measure};

\draw[line width=1pt,->] (-0,0) to ++(0,-0.5);

\hologram{0}{-0.5}{-90};
\node[right] at (0.35,-0.8) {\small $\ket{0}\bra{-l}$};
\node[left] at (-0.35,-0.8) {\small $\ket{0}\bra{+l}$};

\draw[line width=1pt,->] (-0.6,-1.1) to ++(0,-1+0.5-0.1);
\draw[line width=1pt,->] (0.6,-1.1) to ++(0,-1+0.5-0.1);

\node[below] at (0,-1.1-1+0.5-0.1){$\ket{\Psi_{Pol}}=\frac{1}{\sqrt{2}}\left(\ket{H,\vec{k}_{1}}\ket{V,\vec{k}_{2}}+\ket{V,\vec{k}_{1}}\ket{H,\vec{k}_{2}}\right)$};


\draw[line width=1pt,->] (-0.6,-1.1-1+0.5-0.4-0.1) to ++(0,-1+0.5+0.35-0.2+0.1-0.1);
\node[left] at(-0.6,-1.1-1+0.5-0.4-0.1-1+0.5+0.35-0.15+0.1) {1};
\draw[line width=1pt,->] (0.6,-1.1-1+0.5-0.4-0.1) to ++(0,-1+0.5+0.35-0.2+0.1-0.1);
\node[right] at(0.6,-1.1-1+0.5-0.4-0.1-1+0.5+0.35-0.15+0.1) {2};

\filledboxtwo{0+0.1-0.075}{-4+0.5+1.5-0.35-0.1}{1+0.1+0.075}{-4.5+0.5+1.5-0.35-0.1-0.17};
\node[below] at (0.5+0.1,-4+0.5+1.5-0.35-0.1){\small QWP+HWP};
\node[below] at (0.5+0.1,-4.2+0.5+1.5-0.35-0.1){\small +PBS};
\node[below] at (0.5+0.1,-4.4+0.5+1.5-0.35-0.1){\small detector};
\filledboxtwo{-1-0.1-0.075}{-4+0.5+1.5-0.35-0.1}{0-0.1+0.075}{-4.5+0.5+1.5-0.35-0.1-0.17};
\node[below] at (-0.5-0.1,-4+0.5+1.5-0.35-0.1){\small QWP+HWP};
\node[below] at (-0.5-0.1,-4.2+0.5+1.5-0.35-0.1){\small +PBS};
\node[below] at (-0.5-0.1,-4.4+0.5+1.5-0.35-0.1){\small detector};

\node[below] at (0,-4.5+0.5+1.5-0.35-0.25){measuring coincident counts};

\end{scope}
			
\end{tikzpicture}